\documentclass[11pt]{article}
\usepackage{xspace,amssymb,amsmath,helvet,graphicx}
\begin{document}
\newcommand{\dee}{\,\mbox{d}}
\newcommand{\naive}{na\"{\i}ve }
\newcommand{\eg}{e.g.\xspace}
\newcommand{\ie}{i.e.\xspace}
\newcommand{\pdf}{pdf.\xspace}
\newcommand{\etc}{etc.\@\xspace}
\newcommand{\PhD}{Ph.D.\xspace}
\newcommand{\MSc}{M.Sc.\xspace}
\newcommand{\BA}{B.A.\xspace}
\newcommand{\MA}{M.A.\xspace}
\newcommand{\role}{r\^{o}le}
\newcommand{\signoff}{\hspace*{\fill} Rose Baker \today}
\newenvironment{entry}[1]%
{\begin{list}{}{\renewcommand{\makelabel}[1]{\textsf{##1:}\hfil}%
\settowidth{\labelwidth}{\textsf{#1:}}%
\setlength{\leftmargin}{\labelwidth}
\addtolength{\leftmargin}{\labelsep}
\setlength{\itemindent}{0pt}
}}%
{\end{list}}
\title{Reactive Social distancing in a SIR model of epidemics such as COVID-19}
\author{Rose Baker\\School of Business\\University of Salford, UK}
\maketitle
\begin{abstract}
A model of reactive social distancing in epidemics is proposed, in which the infection rate changes with the number infected.
The final-size equation for the total number that the epidemic will infect can be derived analytically,
as can the peak infection proportion. This model could assist planners during for example the COVID-19 pandemic.
\end{abstract}
\section*{Keywords}
Epidemics; SIR model; social distancing; COVID-19
\section{Introduction}

Let the proportions of a population who are susceptible, infected and removed or recovered (now immune) be $s(t), i(t), r(t)$ respectively.
Then the SIR compartmental model can be written:
\[\dee s(t)/\dee t=-\beta s(t)i(t),\]
\[\dee i(t)/\dee t=\beta s(t)i(t)-\alpha i(t),\]
\[\dee r(t)/\dee t=\alpha i(t),\]
where $t$ denotes time, $\beta$ is the rate at which an infected individual infects others and $\alpha$ the rate of recovery/death. The last equation is not needed
as $s(t)+i(t)+r(t)=1$. The basic reproduction number $R_0=\beta/\alpha$.

\begin{figure}
\centering
\makebox{\includegraphics{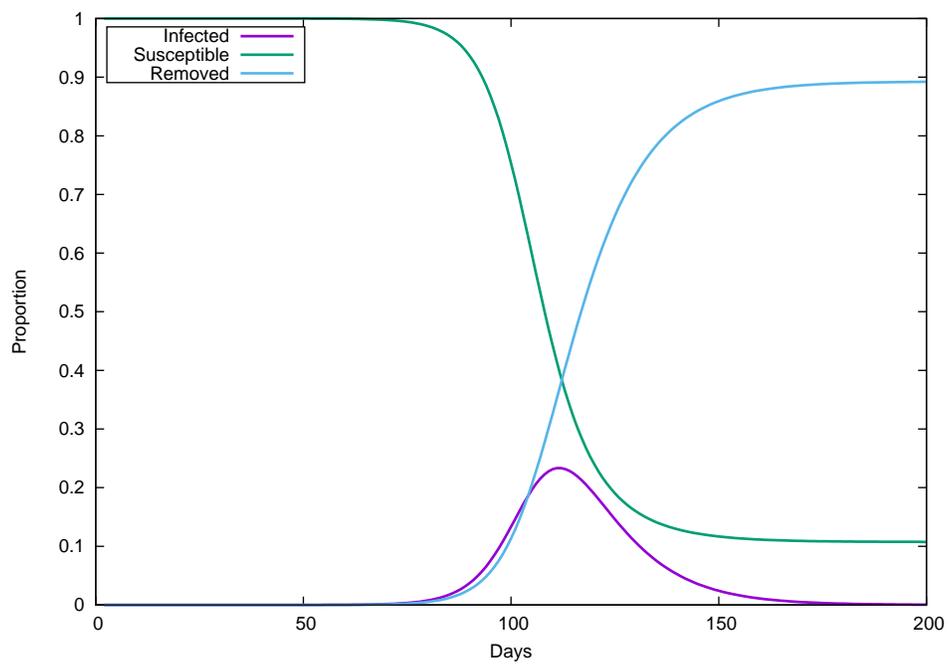}}
\caption{\label{figa}Proportions susceptible, infected and removed under the SIR model when $R_0=2.5$, $\alpha=1/9$ per day, starting with one infected individual in a population of 66.44 million.}
\end{figure}

The properties of this simple model are well understood; see \eg Brauer, Castillo-Chavez and Feng (2019).
Predictions starting with one infected case are shown in figure \ref{figa}.

This model is simplistic, \eg because it assumes exponentially-distributed dwell times in the compartments and ignores differences in behaviour arising from demographics.
However, this model is more than a `toy' model, and can be used to make quite useful predictions of the course of
of epidemics. For example, for SIR models, a final size relation can be derived, which is a nonlinear equation for $s_\infty$, the proportion
who remain uninfected when the epidemic is over. Starting with a susceptible population, the formula is
\[-\ln(s_\infty)=R_0(1-s_\infty).\]

A feature of diseases such as influenza and those caused by coronaviruses such as SARS, MERS and COVID-19, is that individuals practice social distancing
to reduce the contact rate $\beta$. If $R_0$ could be reduced below unity, an epidemic would fade away and could be got rid of, but except in a small isolated community some virus would always remain, and
when social distancing was relaxed, the epidemic would begin anew. However, as the rate of growth of the number infected is $(R_0-1)\alpha$, decreasing $R_0$
slows the epidemic, and reduces the percentage infected. 

In this article a simple model of social distancing is introduced. Yu {\em et al} (2017) describe such models. What is novel here is that the proportion infected $i(t)$ can be found in terms of $s(t)$,
and so a final-size equation can be derived, plus the peak proportion of the population infected. Such formulae may help planners trying to slow the epidemic to reduce peak demand on health services
while preventing the total shutdown of industry.
\section{The model}
The contact rate $\beta$ is replaced by $\beta(t)=\beta/(1+\gamma i(t))$. Hence the contact rate decreases as infection levels rise, and increases again as they fall.
As infection level $i(t)$ tends to zero, we regain the regular SIR model, so $R_0$ is unchanged. 

The function $i(t)/(1+\gamma i(t))$ is a Hill function. Many other functions could be used here, but with the Hill function an analytic solution is possible.

The new model is thus:
\[\dee s(t)/\dee t=-\frac{\beta s(t)i(t)}{1+\gamma i(t)},\]
\[\dee i(t)/\dee t=\frac{\beta s(t)i(t)}{1+\gamma i(t)}-\alpha i(t).\]

\begin{figure}
\centering
\makebox{\includegraphics{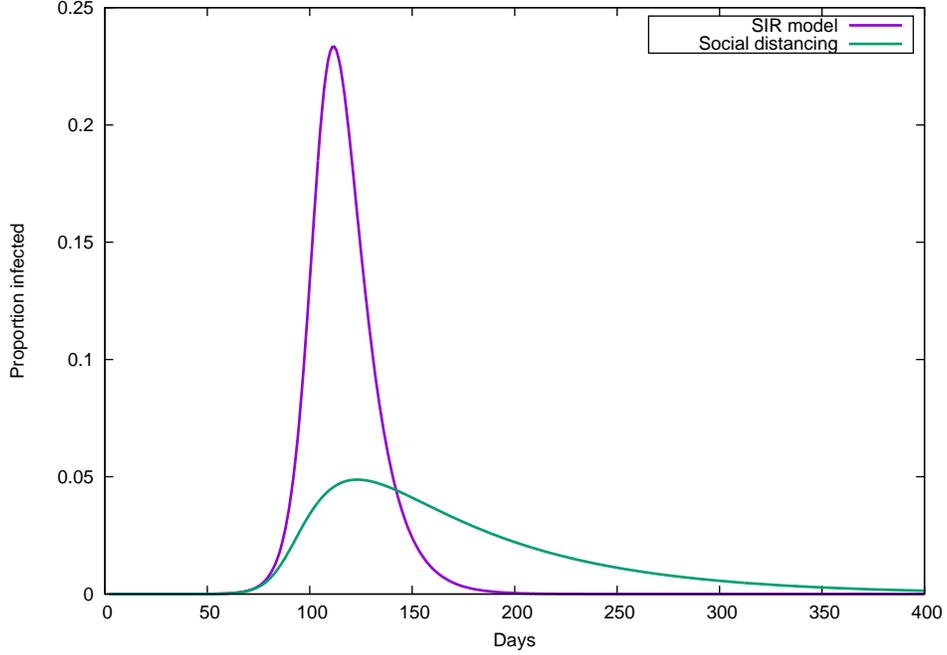}}
\caption{\label{figb}Proportions infected under the standard SIR model and with $\gamma=20$ under the social distancing model when $R_0=2.5$.}
\end{figure}

The model is a reasonable model of behaviour, because numbers dying daily will be widely known, which is a slightly-lagged proxy for $i(t)$.
Hence individuals have the information on which to act. Government-mediated social distancing may also approximate the situation of reactive distancing.
Figure \ref{figb} shows the evolution of proportions infected for the model with $\gamma=0$ and $\gamma=20$.

Dividing the model equations, we have
\begin{equation}\dee i(s)/\dee s=-\frac{(\beta s-\alpha)/(1+\gamma i(s))-\alpha}{\beta s/(1+\gamma i(s))},\label{eq:diff}\end{equation}
where a factor of $i(s)$ has been cancelled out.
Rearranging,
\[s\dee i(s)/\dee s-\frac{\alpha\gamma}{\beta}i(s)=\alpha/\beta-s.\]
Writing $x=\ln(s)$, we have that
\[\dee i(x)/\dee x-\frac{\alpha\gamma}{\beta}i(x)=\alpha/\beta-\exp(x).\]
This is an ordinary differential equation for $i(x)$, which can be solved by standard methods.
These are: solve the homogeneous equation to get the complementary function, find the particular integral
using for example the D-operator method, and find the unknown constant multiplying the complementary function by requiring that $i=0$ when $s=1$.
This yields
\begin{equation}i(t)=(\frac{1}{\gamma}+\frac{1}{1-\alpha\gamma/\beta})s(t)^{\alpha\gamma/\beta}-\frac{1}{\gamma}-\frac{s(t)}{1-\alpha\gamma/\beta},\label{eq:i}\end{equation}
where we have transformed back from $x$ to $s$.

When $\gamma=R_0$ the solution method fails, and the equation becomes
\[\dee i/\dee x-i=1/R_0-\exp(x).\]
Solving this, we obtain 
\begin{equation}i(s)=(s-1)/R_0-s\ln s.\label{eq:spec}\end{equation}

In general, the final-size equation for $s_\infty$, the number never infected, can now be found by setting $i(t)=0$.
Then
\begin{equation}s_\infty^{\gamma/R_0}=\frac{1/\gamma+s_\infty/(1-\gamma/R_0)}{1/\gamma+1/(1-\gamma/R_0)}.\label{eq:sinf}\end{equation}
Equation (\ref{eq:sinf}) can be solved numerically for $s_\infty$ using for example the Newton-Raphson method of iteration, and converges quickly.
For the special case, $(s-1)/R_0=s \ln s$.

Also of interest is the peak proportion infected. Setting $\dee i/\dee s=0$ in (\ref{eq:diff}) yields $i_{\text{max}}=\frac{R_0}{\gamma}(s_{\text{max}}-1/R_0)$,
and substituting in (\ref{eq:i}) gives the nonlinear equation
\[s_{\text{max}}=\frac{(1+\gamma-\gamma/R_0)}{R_0}s_{\text{max}}^{\gamma/R_0}.\]

This can be solved on taking logarithms, when
\[s_{\text{max}}=(\frac{1-\gamma/R_0+\gamma}{R_0})^{\frac{1}{1-\gamma/R_0}},\]
\begin{equation}i_{\text{max}}=(\frac{R_0}{\gamma})(\frac{1-\gamma/R_0+\gamma}{R_0})^{\frac{1}{1-\gamma/R_0}}-1/\gamma.\label{eq:imax}\end{equation}
As $\gamma\rightarrow 0$ we regain $s_{\text{max}}=1/R_0$.
As $\gamma\rightarrow\infty$, the left-hand side of (\ref{eq:sinf}) goes to zero as $s_\infty < 1$. Then $s_\infty \rightarrow 1/R_0$,
the minimum number to give herd immunity and bring the epidemic to an end. Thus extreme reactive social distancing gives the minimum number of infections possible,
whilst necessarily prolonging the duration of the epidemic.

\begin{figure}
\centering
\makebox{\includegraphics{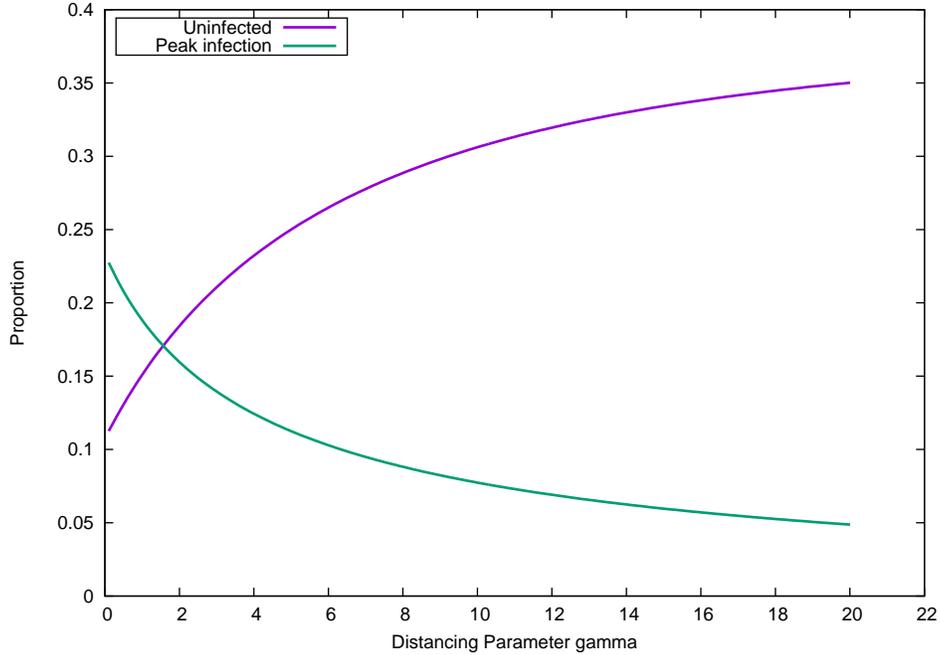}}
\caption{\label{figc}Proportion remaining uninfected and peak proportion infected under the social distancing model when $R_0=2.5$.}
\end{figure}

For the special case, $s=\exp(-(1-1/R_0))$ and $i_{\text{max}}=s-1/R_0$.
These special-case solutions can be found either from (\ref{eq:spec}) or by using L'Hospital's rule.
Figure \ref{figc} shows $s_\infty$ and $i_{\text{max}}$ plotted against $\gamma$.

Finally, from (\ref{eq:imax}), as $\gamma\rightarrow\infty$, the effective reproduction number at peak infection $R_0/(1+\gamma i_{\text{max}})\rightarrow 1$.

\section{Conclusions}
An epidemic model with reactive social distancing has been described, for which the final number of infections and the peak infection proportion can 
be found analytically. Unlike a fixed reduction in contact rate, reactive social distancing allows the epidemic to complete and not restart when distancing is relaxed.
In the limit that $\gamma\rightarrow\infty$ the number left uninfected decreases to $1/R_0$.

How could this model be used as a planning aid when the value of $\gamma$ is unknown? The model or a more sophisticated one could be formally or informally fitted to data
and the effective value of $\gamma$ estimated. That would give the likely course of the epidemic. One could then see how much more (or less) social distancing would be required for the optimal outcome
that balances mortality and pressure on health services against economic disruption.


\begin{thebibliography}{99}
\bibitem{brauer}Brauer F., Castillo-Chavez, C. and Feng Z. (2019) Mathematical methods in Epidemiology, Springer, new York.
\bibitem{yu}Yu D., Lin Q., Chiu A. P. Y., He D. (2017) Effects of reactive social distancing on the 1918 influenza pandemic. PLoS ONE 12(7): e0180545.
\end{thebibliography}
\end{document}